\documentstyle[12pt]{article}
\begin{document}

\title{Nonlinear Liquid Drop Model. Cnoidal Waves}
\author{Andrei Ludu$^1$ , Aureliu S\u{a}ndulescu$^2$ and Walter Greiner \\
\normalsize{Institut f\"ur Theoretische Physik der J.W.Goethe
Universit\"at, } \\
\normalsize{ D-60054 Frankfurt am Main, Germany}}
\date{}
\maketitle
\begin{abstract}
By introducing in the hydrodynamic model, i.e. in the hydrodynamic equations
and the corresponding boundary conditions, the higher order terms in the
deviation of the shape, we obtain in the second order the Korteweg de Vries
equation (KdV). The same equation is obtained by introducing in the liquid drop model
(LDM), i.e. in the kinetic, surface and Coulomb terms, the higher terms in
the second order. The KdV equation has the cnoidal waves as steady-state 
solutions. These waves could describe the small anharmonic
vibrations of spherical nuclei up to the solitary waves.
The solitons could describe the preformation of
clusters on the nuclear surface. 
We apply this nonlinear liquid drop model to the alpha 
formation in heavy nuclei. We find an additional minimum in the
total energy of such systems, corresponding to the solitons as clusters
on the nuclear surface. By introducing the shell effects we choose this
minimum to be degenerated with the ground state. 
The spectroscopic factor is given by the ratio of the square
amplitudes in the two minima.
\end{abstract}

\setlength{\baselineskip} {4ex.}
\footnotetext[1]{Permanent Address: Bucharest University,
Bucharest-Magurele, PO.Box MG-5211, Romania; 
e-mail:ludu@th.physik.uni-frankfurt.de}
\footnotetext[2]{Permanent Address:Institute of Atomic Physics, 
Department of Theoretical
Physics, Bucharest-Magurele, P.O.Box MG-6, Romania; 
e-mail:sandulescu@roifa.ifa.ro and Romanian Academy, Calea Victoriei 125,
71102 Bucharest, Romania; E-mail: aursand@aix.acad.ro}
\vfill
\eject

\section{Introduction}
\vskip 1cm

It is well known that liquid drop model, as a collective model of the
nucleus describes excelently the spectra of spherical nuclei as small
vibrations (harmonic in the linear approximation or anharmonic in higher
approximations) around their shape.  On the other
hand it is known that on the nuclear surface of heavy nuclei close to the
magic nuclei ($^{208}$Pb, $^{100}$Sn) a 
large enhancement of clusters (alpha,
carbon, oxigen, neon, magnesium, silicon) exists which leads to 
the emission of such clusters as natural decays [1,2].
It is also clear that traditional collective models [3] are not able to give a
complete explanation of such natural decays, i.e. they still did not
completely answer
to the main physical question: why should nucleons join together and
spontaneous form an isolated cluster on the nuclear surface ? 
Only by the introduction in the shell model of many body correlation effects 
we could form an isolated bump, stable in time. It is possible to describe
the formation of such clusters in a collective model [4,5] ? 

In the present paper by introducing the nonlinearities in the liquid
drop model we succeded to give a positive answer to this problem.
 In a nonlinear liquid drop model we can describe simultaneously, by 
cnoidal waves the transitions from small vibrations to the 
formation of solitons.
The experimental discovery and the theoretical foundation of solitons [6] 
as non-dispersive localized waves moving
uniformly, lead to a powerfull theory of classical
field equations with such solutions [7-9]. Also it was possible to explain the
correspondence between classical soliton solutions and the extended-particle
states of the quantized version of the theory.
This lead to a generalisation of the semiclassical expansion of
quantum mechanics in quantum field theory.
In the last case solitons (and breathers or instanton solutions) are
non-perturbative. Of course, the above methods work only if the initial physical
model allows first the existence of some non-trivial classical localized
solutions.

From the mathematical physics point of view solitons, as solutions of 
non-linear
evolution equations, are isolated waves which preserve their shape and have
finite and localized energy density. Recently has been shown that solitons 
may be also relevant in nuclear physics [4,5] or particle 
physics [10].
Also it was realized that many field theoretical models for
particle interactions, and even for quantum extended particles [11],
possess soliton or breather solutions and that the solitons ought to be
interpreted as additional particle-like structures in theory.
The traveling solutions of
the KdV equation [7] and singular
solutions having poles at $\pm \infty$.
Recently, in order to describe the quasimolecular spectra, we have introduced 
in [5] a one-dimensional soliton model 
for the cluster (alpha particle) and the rotator-vibrator model for the
nucleus. An excelent agreement with the experimental data was obtained.

We conclude that the nonlinear terms lead to new qualitative picture
of the liquid drop model, i.e. the harmonic oscillations grow
into anharmonic ones which can lead to a stable soliton configuration.
Both the potential picture and the phase space portraits support this
behaviour. We stress that all these results are embedded into a Hamiltonian
formalism.

In the present paper we first introduce in the hydrodynamical model the
higher terms in the deviation of the spherical shape. We have shown
that in the third order we obtain the Korteweg de Vries equation (KdV).
In the following chapter 3 by introducing in the liquid drop model the third
order
terms in the deviation of the shape we obtained the same KdV equation
as in the nonlinear hydrodynamical model. We should like to stress that
these nonlinear equations are Hamiltonian eqautions which describe the total
energy of the system. Chapter 4 contains the nonlinear solutions: cnoidal
waves
and the singular solutions for KdV equation.
Last chapter discuss the application of nonlinear liquid drop
model to the alpha preformation factors.

\vskip 1cm
\section{The nonlinear hydrodynamic  model}
\vskip 1cm

There are two possible ways to describe the classical dynamics of a
liquid drop: first is the fluid (hydrodynamic) approach based on the 
continuity, Euler and the
equations of state together with boundary conditions and
the second is the Hamiltonian approach. However there is a deep connection between 
these two approaches.
The motion of a perfect incompressible liquid in a domain is governed 
only by the Euler equation, since the continuity and state equations 
reduce in this case to the Poisson equation.
The boundary conditions ask the inner product between the velocity field and 
the volume differentiable 3-form to be zero (special procedure occures when the boundary 
itself is variable in time - free surface - and results in nonlinear 
contributions to
the differential equations). The flow of the velocity field belongs to
the group of volume preserving transformations of the volume of the drop and
hence it could be a geodesic on the manifold of this group.
Now, the connection with the second approach comes from the observation that
the Euler equation is Hamiltonian in the sense that, between all other possible
flows, only those which satisfy  Euler equation are geodesics.
In the case of the liquid drop model the dominant terms are the volume, the
surface and Coulombian.
All terms are dependent on the geometry of
the surface. Therefore, we expect that the dynamics of the liquid drop is
strongly dominated (phenomenologicaly) by the dynamics of the free
surface, in both these approaches. 
 
Let us describe the surface of the nucleus in spherical coordinates 
$(r,\theta ,\phi )$ as a
function of the polar angles $\theta $ and $\phi $, by writing the nuclear
radius in the form
\begin{eqnarray}
r=R_0 (1+\xi (\theta , \phi ,t)),
\end{eqnarray}
where $R_0$ is the radius of the spherical nucleus and 
the shape function $\xi $ is the
difference in the radius between the deformed and the spherical one.
In the following we search for deformed surfaces which contain   
stable traveling waves which can lead to a bump.
The stability is fulfilled if the
function describing the surface arises as a solitary wave solution of one of 
the classical nonlinear equations like KdV (Korteweg de Vries),
whose stability was clearly established in literature
[7,12,13]. These equations are tractable as dynamical systems
in the frame of nonlinear infinite-dimensional Hamiltonian 
field theories. 
The cnoidal waves as solutions of KdV equation describe small vibrations up
to the solitary waves. In the case of the spherical surface, the
localization condition is realised for values of the angular half-width $L$ 
of the shape function $\xi (\theta ,\phi ,t)$ smaller than $\pi $, at any 
moment of time [4,5]. 
Without any loss of generality and in the spirit of the above theory,
it is convenient to look for a special 
space-time behaviour of the shape function  of the form 
\begin{eqnarray}
\xi (\theta ,\phi ,t)=g(\theta )\eta (\phi -Vt)
\end{eqnarray}
with $g$ an arbitrary bounded, non-vanishing continuous function, 
$\eta $ a compact supported, or rapidly
decreasing function with $V$ defining the tangential
velocity of the traveling solution $\eta $ on the surface. 
This solution represents a stable traveling perturbation ($\eta $)
in the $\phi $ direction, having a given transversal profile ($g$)
in the $\theta $ direction.
This is  different from the traditional liquid drop model case
when one expands the shape function  in spherical harmonics and 
where we have shown [5] one needs to consider more than 10 multipoles to fit 
such shapes of a localised bump.

We suppose (without any
loose of generality, due to the spherical symmetry) that the bumps are 
situated on a circle $\theta =\pi / 2 $
such that the variable $\theta $ plays only the role of a parameter in the
corresponding dynamical equations, and that the bumps travel
along the $\phi $ coordinate only.
This choice results in the separation of variables in the shape function 
in eq.(2).
Since in the investigation of the dynamics of the surface the coordinate $r$
will be involved only in the (free surface) boundary conditions, then $\phi $
remains the unique free coordinate. Hence, we can reduce the whole 
3-dimensional problem to a 1-dimensional formalism.

In the hydrodynamic approach we treat the nucleus is a perfect ideal fluid 
layer (incompressible,
irrotational and without viscosity) described by the field velocity
${\vec v}(r,\theta ,\phi ,t)$ and by the constant mass density $\rho =$const. 
From the continuity 
equation we have $div $ ${\vec v}=0$ and due to the irrotationality
condition, $\nabla \times {\vec v}=0$, we have a potential flow, described by
the velocity potential $\Phi (r,\theta ,\phi ,t)$,
and the corresponding
Laplace equation 
\begin{eqnarray}
{\vec v}=\nabla \Phi, \ \ \ \ \ \triangle \Phi =0.
\end{eqnarray}
The dynamics of this perfect fluid is described by the Euler equation
\begin{eqnarray}
{{\partial {\vec v}} \over {\partial t}}+({\vec v}\cdot \nabla  ){\vec v}
=-{{1} \over {\rho }}\nabla P+{{1} \over {\rho }}{\vec f},
\end{eqnarray}
where $P$ is the pressure and ${\vec f}$ is the volume density of the forces
acting into this fluid, e.g. for the Coulombian one we have $
{\vec f} =-{\rho }_{el}\nabla \Psi$, with $\Psi $ the electrostatic
potential and $\rho _{el}$ the charge density, supposed to be constant, too. 
By using eqs.(3), eq.(4) becomes, in the Coulombian case 
\begin{eqnarray}
\biggl ( \Phi _{t}+{1 \over 2}|\nabla \Phi |^{2}\biggr ) \bigg | _{\Sigma }
=-{{1} \over {\rho }}P
-{{\rho _{el}} \over {\rho }}\Psi |_{\Sigma }.
\end{eqnarray}
To determine uniquely the unknown functions $\Phi $ and $\xi $
we need, in completion of eqs.(3,5), the boundary conditions for the scalar
harmonic field $\Phi $, on (maximum) two closed surfaces: the external 
free surface of the nucleus described by eqs.(1) or (2) 
and the inner surface (if it exists) of the fluid layer. 
The latter condition can be expressed in a simpler 
form if we consider that the motion is limited to only
a thin fluid layer characterised by zero radial velocity of the flow
on its inner surface. 
This last condition expresses the existence of a rigid core
in the volume of the nucleus.
The first boundary condition can be expressed in the most general form
of the kinematical constrain of the free surface of the fluid 
described by eq.(1), [4,5,7] 
\begin{eqnarray}
{{dr} \over {dt}} \biggr | _{\Sigma }=
{\biggl ( {{\partial r} \over {\partial t}}+
{{\partial r} \over {\partial \theta}}{{d\theta } \over {dt}}+ 
{{\partial r} \over {\partial \phi}}{{d\phi } \over {dt}}
\biggr ) } \biggr | _{\Sigma },
\end{eqnarray}
where $r(t,\phi ,\theta )$ in the RHS represents the shape function described in eq.(1) 
and the label $\Sigma $ means that eq.(6) is taken on the free surface 
$\Sigma $. 
This equation allows very general types of movements,
including traveling and vibrational waves. Eq.(6) reduces 
to the form ${{dr} \over {dt}}\biggr | _{\Sigma }=
{{\partial r} \over {\partial t}} \biggr | _{\Sigma }
$ when one considers only its linear approximation [3,14], 
i.e. that one used in the Bohr-Mottelson model. This linearization
implies the existence of only collective radial vibrations and does no 
allow any motion along the tangential direction. Eq.(6) 
can be written  in terms of the 
derivatives of the potential of the flow and the shape function $\xi $
\begin{eqnarray}
\Phi _{r} \biggr | _{\Sigma }= R_0 {\biggl ( 
\xi _{t} +
{{\xi _{\theta }} \over {r^2}}\Phi _ {\theta }+
{{\xi _{\phi }} \over {r^2 \sin ^{2} \theta }}\Phi _{\phi }
\biggr ) } \biggr | _{\Sigma }, 
\end{eqnarray}
where ${{\partial \Phi } \over {\partial r}}=v_r=\dot r$ is the radial
velocity and ${{1} \over {r}}
{{\partial \Phi } \over {\partial \theta}}=v_{\theta }=r\dot{\theta } $,
${{1} \over {r \sin \theta}}
{{\partial \Phi } \over {\partial \phi }}=v_{\phi }=r \dot {\phi }\sin
\theta$ 
are the tangential velocities. We denote here the partial differentiation
by suffixes, $\partial \Phi /\partial \phi =\Phi _{\phi }$, etc.
The existence of a rigid core of radius $R_0 -h(\theta )>0$,
$h(\theta)\ll R_0 $, introduces the second boundary condition for the radial velocity 
on the surface of this core in the form
\begin{eqnarray}
v_r |_{r=R_0 -h}={{\partial \Phi } \over {\partial r}} \biggr | _{r=R_0 -h}=0.
\end{eqnarray}
Both eqs.(7,8) are von Neumann type of boundary conditions.
The motion of the fluid is described by the Laplace equation
eq.(3) for $\Phi $, and by the two boundary conditions, eqs.(7,8), for
$\Phi $ and $\xi $. To these equations we have to add the dynamical
equation in the form of Euler-Lagrange equation if we use a Lagrangean
formalism, or in the form of Hamilton equation if we use a Hamiltonian
formalism.
In the present paper we consider the contribution of the nonlinear terms
in all equations (e.g. the second and third terms of RHS of eqs.(6,7)). The
corresponding solutions should reduce to the standard normal modes of 
vibrations, if we restrict to the linear approximation.

For such typical hydrodynamical problems, like that described by
eqs.(3,5,7,8), one generaly uses, in the linear approximation, the expansion
in spherical harmonics. Such an expansion is no more 
appropriate 
for the nonlinear cases. Hence, we use for the potential of the flow the expansion 
\begin{eqnarray}
\Phi=\sum_{n=0}^{\infty}\biggl (
{{r-R_0 } \over {R_0 }}
\biggr )^{n}f_{n}(\theta , \phi ,t),
\end{eqnarray}
where the functions $f_n$ are not orthogonal on the surface of the sphere
and do not form in general a complete system. The convergence of eq.(9)
is controlled by the value of the small quantity ${{r-R_0 } \over {R_0 }}
\le max|\xi |=\epsilon$, [4].
From the Laplace equation (in spherical coordinates), and the expansions 
\begin{eqnarray}
{{1} \over {r^n }}={1 \over {R_{0}^{n} }}\sum_{k=0}^{\infty }
(-1)^k ((n-1)k+1)\xi ^{k}, \ \ \  \ k=1,2,
\end{eqnarray}
we obtain a system of equations which result in the recurrence
relations for the unknown functions $f_n $
\begin{eqnarray}
f_n= {{(-1)^{n-1}(n-1)\triangle _{\Omega }f_{0}-2(n-1)f_{n-1}+
\sum_{k=1}^{n-2}(-1)^{n-k}(2k-(n-k-1)\triangle _{\Omega} f_{k})} 
\over {n(n-1)}},
\end{eqnarray}
with $n\ge 2$ and where $\triangle _{\Omega }=
{{1} \over {\sin \theta }}{{\partial } \over {\partial \theta}}
\biggl ( \sin \theta {{\partial } \over {\partial \theta}}
\biggr ) +{{1} \over {\sin ^{2} \theta }}{{\partial } \over {\partial \phi }}
$ is the angular part of the
Laplacean operator in spherical coordinates. 
Eq. (11) reduces the unknown functions to only two: 
$\triangle _{\Omega }f_0 $ and $f_1 $:
$$
f_2 =-{1 \over 2}(\triangle _{\Omega }f_0 +2f_1 ),
$$
\begin{eqnarray}
f_3 ={1 \over 6}(4\triangle _{\Omega }f_0 -
4\triangle _{\Omega }f_1 +4f_1 +2),
\end{eqnarray}
$$
f_4 ={1 \over 24}({\triangle }^{2}_{\Omega }f_0 -
14\triangle _{\Omega }f_0 +8\triangle _{\Omega }f_1 -8f_1 ) \ \  \dots .
$$
If we choose the independent functions $\triangle _{\Omega }f_0 $ and $f_1 $
to be smooth on the sphere, they must be bounded together with all
the $f_n $'s (these being linear combinations of higher derivatives of
$f_{0}$ and $f_1 $) and hence the series in eq.(9) are indeed controlled by  the
difference in the radius between the deformed and the spherical one.
However, in the following we will use only truncated polynomials of these
series. 

By introducing eqs.(11-12) in the second boundary condition, eq.(8), we obtain
the condition
\begin{eqnarray}
\sum_{n=1}^{\infty }n\biggl ( -{{h} \over {R_0 }}
\biggr )^{n-1}f_n =0,
\end{eqnarray}
which reads, in the first order in $h/R_0 $
\begin{eqnarray}
f_1 ={{2h} \over R_0 }f_2 .
\end{eqnarray}
From eqs.(12,14) the unknown function $f_1 $ is obtained,
in the smallest order in $h/R_0 $
\begin{eqnarray}
\triangle _{\Omega }f_0 =-\biggl ( {{R_0 } \over {h}}+2
\biggr ) f_1 .
\end{eqnarray}
Concerning the first boundary condition held at the free surface $\Sigma $,
eq.(7), we need to calculate the derivatives of the potential of the flow on that
surface 
$$
\Phi _{r}|_{\Sigma }=\sum_{n}n{{(r-R_0 )^{n-1}_{\Sigma }} \over {R_{0}^{n}}}
f_{n}
= {{f_{1}} \over {R_0 }}+{{2\xi f_2 } \over {R_{0}}}+{\cal O}_{2}(\xi ),
$$
\begin{eqnarray}
\Phi _{\phi }|_{\Sigma }=\sum_{n}\xi ^{n}f_{n,\phi }=
f_{0,\phi }+\xi f_{1,\phi }+{\cal O}_{2}(\xi ),
\end{eqnarray}
$$
\Phi _{\theta }|_{\Sigma }=\sum_{n}\xi ^{n}f_{n,\theta }=
f_{0,\theta }+\xi f_{1,\theta }+{\cal O}_{2}(\xi ).
$$
By introducing the series eqs.(10,16) in eq.(7), for the traveling wave
solution eq.(2), we have the equation
\begin{eqnarray}
f_1 +2\xi f_2 =
R_{0}^{2} \xi _t+ {{\xi _{\phi }(1-2\xi )} \over {\sin ^{2}\theta }}
(f_{0,\phi }+\xi f_{1,\phi})+
\xi _{\theta }(1-2\xi )(f_{0,\theta }+\xi f_{1,\theta }) .
\end{eqnarray}
We keep the nonlinearity of the boundary conditions, eqs.(7,8,13), in the first
order (i.e. the first order in the expression of $f_0$ and the second order 
in the expression of $f_1$). Consequently, in order to be consistent, it is
enough  to take the linear approximation of
the solution for $f_1 $ in eq.(17), like in the case
of the normal modes of vibrations
\begin{eqnarray}
f_1 =R_{0}^{2}\xi_{t}+{\cal O}_{2}(\xi ).
\end{eqnarray}
Hence, by introducing the linear approximation for $f_1 $
(eq.(18)) in eq.(17)
\begin{eqnarray}
2\xi f_2 =
{1 \over {\sin ^2 \theta }}\biggl (
-\xi _{\phi }f_{0,\phi }
+\xi \xi _{\phi }(f_{1,\phi }-2f_{0,\phi }) \biggr )
+\xi \xi _{\theta }(f_{1,\theta }-2f_{0,\theta }),
\end{eqnarray}
and by taking the expression of $f_2 $ from the recurrence relations,
eq.(14) and  $\triangle _{\Omega }f_0 $ from eq.(15), we obtain the form of
$f_0 $, in the second order in $\xi $
\begin{eqnarray}
f_{0,\phi}=-{{R_{0}^{3}\sin ^2 \theta } \over {h}}{{\xi \xi _{t}} \over {\xi
_{\phi}}}(1+2\xi )-{{\xi _{\theta}f_{0,\theta }} \over {\xi _{\phi }}}
+{\cal O}_{3}(\xi ).
\end{eqnarray}
In the case of traveling wave profile of the form
$\xi (\theta ,\phi ,t)=g(\theta )\eta (\phi -Vt)$, which
introduces the restriction $\xi _{\phi  }=-V \xi _{t}$ and vanishes the
tangential velocity in the $\theta $-direction,
eq.(20) becomes 
\begin{eqnarray}
f_{0,\phi }={{VR_{0}^{3}\sin ^{2}\theta } \over {h}}\xi(1+2\xi )
+{\cal O}_{3}(\xi ).
\end{eqnarray}
Eqs.(18,20,21) describe, in the second order in $\xi$, the connection
between the velocity potential (the flow) and the shape function, through
the boundary conditions. This fact is a typical feature of nonlinear systems. 
The dependence of $\Phi |_{\Sigma }$ on the polar angles, in the second
order in $\xi $, has the form of a
quadrupole in the $\theta $-direction and depends only on $\xi $ and its
derivatives in the $\phi $-direction. For traveling wave profiles
the tangential velocity in the direction of motion of the perturbation,
$v_{\phi }=\Phi _{\phi }/r\sin \theta $ is proportional with $\xi $
in the first order
\begin{eqnarray}
v_{\phi }={{2VR_{0}\sin \theta } \over {h}}\xi+{\cal O}_{2}(\xi ),
\end{eqnarray}
with the constant of proportionality being exactly the coupling parameter
$\chi $ from our previous solitonic model, i.e. eq.(20) in [4]. This can be
seen also from Figs. 1 and 2a in [4], where it is clearly stated the $1/h$
dependence of $\chi $, for fixed velocity $V$, like in eqs.(21,22).  
Eqs.(20-22) are valid for any traveling wave shape functions, case
which includes  harmonic or  anharmonic oscillations, solitons,
breathers, cnoidal waves, etc. [7-9,13].

In order to obtain the dynamical equation for the surface $\Sigma $
we follow the usual formalism for the normal vibration of a sphere [14],
corrected with the corresponding nonlinear terms [4,5,7,8], i.e.
we solve the Euler equation on the free surface with respect to the potential
flow $\Phi $ and the shape function $\xi $, for given pressure,
force fields and boundary conditions.
The pressure at the free surface $\Sigma $ can be obtained from the 
surface energy of the deformed nucleus, $U_S$
\begin{eqnarray}
U_S =\sigma R_{0}^{2}\int _{0}^{2\pi }\int _{0}^{\pi}
(1+\xi )\sqrt{(1+\xi )^2 +\xi _{\theta }^{2}
+{{\xi _{\phi }^{2}} \over {\sin ^{2} \theta }}}\sin \theta d\theta d\phi ,  
\end{eqnarray}
where $\sigma $ is the pressure surface coefficient. 
Indeed, by expanding in series the square root in eq.(23) with respect to 
$\xi $, up to the third order, we obtain for the first variation of the functional
$U_S$ 
\begin{eqnarray}
\delta U_S =\sigma
R_{0}^{2}\int_{0}^{2\pi }\int _{0}^{\pi }
\biggl (
2+2\xi +\xi ^2 -
\triangle _{\Omega }\xi+
3\xi ^{2}\xi _{\theta } ctg \theta  
\biggr )\delta \xi \sin \theta d\theta d\phi +{\cal O}_{4}(\xi ).
\end{eqnarray}
Following [14]
the surface pressure on $\Sigma $ is given by the local curvature radius
of the surface,  and from the volume conservation we have
\begin{eqnarray}
P|_{\Sigma }=\sigma \biggl ( {1 \over {R_1 }}+{1 \over {R_2 }} \biggr )=
\sigma R_{0}^{2}{{\delta a_{\xi}} \over {R_{0}^{3}(1+2\xi)}},
\end{eqnarray}
where $R_{1,2}$ are the
principal radii of curvature of the surface of the fluid and $\delta a_{\xi }$
is the term in the paranthesis in eq.(24).
By introducing eq.(24) in eq.(25) we obtain the expression
of the surface pressure as a function of the shape function
\begin{eqnarray}
P|_{\Sigma }={{\sigma } \over {R_0 }}
(-2\xi-4\xi ^2 -\triangle _{\Omega }\xi +3\xi \xi ^{2}_{\theta }
ctg \theta )+const.
\end{eqnarray}
The terms of order three in $\xi _{\phi ,\theta}, \xi _{\phi ,\phi }$
and $\xi _{\theta ,\theta }$, 
could be neglected in eq.(25)
due to the high localisation of the solution (the relative amplitude of the
deformation $\epsilon $ is smaller than its angular halh-width $L$, 
$\xi \xi _{\phi \phi }/R_{0}^{2}\simeq \epsilon ^{2}/L^2 \ll 1$, etc.).

The Coulombian potential is  given by a Poisson equation,
$\triangle \Psi =\rho _{el}/\epsilon _{0}$, with $\epsilon _{0}$ the
vacuum dielectric constant. By using the same method like for $\Phi $, [4],
we obtain in the second order for $\xi $, the form
\begin{eqnarray}
\Psi |_{\Sigma }={{\rho _{el} R_{0}^{2}}\over {3\epsilon _{0}}}
\biggl ( 1-\xi -{{\xi ^2 }\over {6}}
\biggr ) .
\end{eqnarray}

In order to write the Euler equation, eq.(5), for the above restrictions, 
we take the surface pressure from eq.(26), the velocity potential from
eqs.(9,14,18,21), and the Coulombian potential from eq.(27) 
 and we obtain, in the second order in $\xi $
and the first order in its derivatives  
$$
\Phi _{t}|_{\Sigma }+{{V^2 R_{0}^{4}\sin ^2 \theta }\over {2h^2 }}\xi ^2 =
{{\sigma }\over {\rho R_0 }}(2\xi +4\xi ^2+\triangle _{\Omega }
\xi -3\xi ^2 \xi _{\theta }ctg \theta )
$$
\begin{eqnarray}  
+{{\rho _{el}^{2} R_{0}^{2}}\over {3\epsilon _{0}\rho }}
\biggl ( \xi +{{\xi ^2 }\over {6}} \biggr ) +const.
\end{eqnarray}
Neglecting all nonlinear terms in this expression, we obtain  
\begin{eqnarray}
\Phi _{t}|_{\Sigma }=-Vf_{0,\phi }=
-{{\sigma }\over {\rho R_0 }}(2\xi +\triangle _{\Omega }
\xi )-
{{\rho _{el}^{2} R_{0}^{2}}\over {3\epsilon _{0}\rho }}\xi
+const.,
\end{eqnarray}
which, together with the linearised eq.(7), 
i.e. $\Phi _{r}|_{\Sigma }=R_0 \xi _{t}$, describes the normal modes of vibration of the liquid
drop in the presence of the Coulombian field with spherical
harmonics solutions for the angular part and complex exponential for the
time dependence. This linear approximation represents
exactly the reduction of the present model 
to the traditional liquid drop model.

In Eq.(28) is a nonlinear partial differential
equation with respect to $\theta $ and $\phi $.
By differentiating it with respect to $\phi $, by using
eqs.(15,18) and by re-ordering the terms, 
we obtain, in the second order
\begin{eqnarray}
A(\theta )\eta _{t }+B(\theta )\eta _{\phi }+C(\theta )g(\theta )\eta \eta _{\phi }+
D(\theta )\eta _{\phi \phi \phi }=0,                         
\end{eqnarray}
which is a  Korteweg de Vries (KdV) equation 
with variable coefficients depending on $\theta $, as a parameter
$$
A={{VR_{0}^{2}(R_0 +2h)\sin ^2 \theta }\over {h}}; \ \ \  
B=-{{\sigma }\over {\rho R_0 }}{{(2g+\triangle
g)} \over {g}}-{{\rho _{el^{2}}R_{0}^2 }\over {3\epsilon _{0}\rho }};  
$$
\begin{eqnarray}
C=8\biggl (
{{V^2R_{0}^{4}\sin ^4 \theta 
} \over {8h^2 }}-{{\sigma } \over {\rho R_{0}}} \biggr ) -
{{\rho _{el}^{2}R_{0}^2 
}\over {9\epsilon _{0}\rho }} ; \ \ \ 
D=-{{\sigma }\over {\rho R_{0}\sin ^2 \theta }}, 
\end{eqnarray}
where by $\triangle g$ we understand only the action over $\theta $.
First we want to make a qualitative analysis
of the solutions of this nonlinear equation. 
The solutions can be clasificated by either using the
algebraic hierachies and topological arguments [7-9,12,13] or by using the phase
space portrait of these solutions [8,12]. The latter way gives a simpler 
picture of the classes of admisible solutions and, in addition, characterize their
periodic and bounded character. 
In order to comment on the traveling wave 
solutions eq.(2), we use the relation $\partial _{t}=-V\partial _{\phi}$,
with $V$ a parameter to be determined. We can then integrate the resulting
equation  once and write it in the form:
$\eta _{tt}=(VA-A_0 )V^2 /C\eta -BV^2 /2C\eta ^2$.
This is the Newton equation of motion for a one-dimensional particle described by the
coordinate $\eta (t)$ in a "potential force"  given by the RHS of the 
above expression. By integrating once this "force" we obtain the
effective one-dimensional potential associated with this motion, $U(\eta
,V)$,
depending on $V$ as a parameter. For a given value of $V$ we can construct
a certain phase space of this type of solutions. 
All possible solutions of the KdV equation are classified according
with all admissible trajectories of this particle, under the action of the 
effective potential, in the phase space, [8,12], Figs.1. 
In Fig.1a we present the effective potential shape and the associate
trajectories in the phase space for the KdV equation. The possible
trajectories with constant energy ($E=U(\eta ,V)+{1 \over 2}\eta _{t}^{2}$), 
are classified through the parameter energy, $E$. 
If we reduce to the linear approximation, the effective potential is
quadratic and the solutions allow only harmonic oscillations, Fig1b.
This situation represents the case of the traditional liquid drop model 
where only the harmonic oscillations are taken into account, and the
governing equation becomes the Helmoltz equation, [3,15].
The KdV equation has a cubic potential energy, Fig1b.
This is an example in which one can see directly the effect of the
introduction of the 
nonlinearity: the harmonic oscillator effective potential gets a pocket and
a saddle point,
which is responsible for the new soliton solutions. Oscillations still exist
and become
anharmonic, being described by the cnoidal wave solutions of the KdV
equation. At the superior boundary of the pocket, where the energy
$E$
equals the potential energy at the saddle, the anharmonic oscillations 
become aperiodic and describe the soliton solution of the KdV equation.
For values of the energy higher than the saddle (the soliton energy), or for 
smaller than the bottom of the valley, the solutions decay into instable,
singular ones, having poles towards infinity. However, these residual
solutions are not taken into account here since they are not traveling
waves.

Eq.(30) has as solutions the cnoidal waves. The localised solution, solitons have the form
\begin{eqnarray}
\eta (\phi ,t)=g(\theta )sech ^2 \biggl ( {{\phi -V(\theta )t} \over {L(\theta )}}
\biggr ),
\end{eqnarray}
where $g(\theta )\le \eta _0 =\epsilon R_0 $ 
is the soliton amplitude, $V(\theta )$ is its angular velocity
and $L(\theta )$ is its half-width. Due to the nonlinearity of the equation, these 
coefficients depend on the coefficients of the KdV equation, 
through the relations, [7,8]
\begin{eqnarray}
L(\theta )=\sqrt{{{12D} \over {\eta _0 C}}}; \ \ \ \ \ V(\theta )=
{{g(\theta) C+3A}\over {3B}},
\end{eqnarray}
in other words, a higher soliton moves faster and has a larger
half-width. 
The soliton is extremely stable against perturbations in the sense of smooth
modifications of the coefficients of the equations, initial conditions,
introduction of small additional terms in the equation or when
interacting with other solutions of this equation (scattering).
Solitons have a constant traveling profile in time and an infinite number of
integrals of motion [7-9,12,13].
Details concerning the explicit dependence of the soliton
parameters, as functions of the coefficients of the KdV equation eq.(30), 
are given {\it {in extenso}} in [4]. 

In the following we analyse the stability of the steady-wave solutions
of the KdV equation, eq.(30), against the dependence of its coefficients
on $\theta $.
In order to have a real traveling wave along the $\phi $-direction, the
parameters of the soliton solution must fulfil the following restrictions:

1. The amplitude of the soliton and the depth of the layer must 
decrease from their maximum values, ($\eta _{0},h$)
on the circle $\theta =\pi /2$ towards 0, when $\theta \rightarrow 0,\pi $. 
Then, the function $g(\theta )$ must increase, when 
$\theta : 0, \pi \rightarrow
\pi /2, $ from $0$ to $\eta _{0}$.  
The function $h(\theta )$ must increase when 
$\theta : 0, \pi \rightarrow
\pi /2, $ from $0$ to $h$.  
The numbers $\eta _{0}, h(\pi /2)$ (i.e. the maximum amplitude of the
soliton and the maximum depth of the layer) are free
parameters, as have been shown in [4,5].

2. The (angular) half-width $L$, eq.(33), must be constant:
\begin{eqnarray}
D(\theta )={\cal C}_1 C(\theta )g(\theta ), 
\end{eqnarray}
3. The (angular) velocity $V$, eq.(33), must be also constant:
\begin{eqnarray}
D(\theta )={\cal C}_2 B(\theta ), 
\end{eqnarray}
where ${\cal C}_{1,2}$ are constants.
Taking into account the mutual relations between the parameter of the
soliton solution and the coefficients of the KdV equation, eqs.(33) and
the restrictions 1-3, we obtain a system
of two differential equations, eqs.(34,35), for the two unknown functions 
$L(\theta ),V(\theta )$,
with bilocal Cauchy conditions at the ends of $[0, \pi ]$, given by
condition 1. Consequently we have for these functions a well defined
boundary condition problem with unique solutions, depending parametricaly
on $\eta _{0},h(\pi /2)$. Both differential equations, eqs.(34,35), are
well defined in $\theta [0,\pi ]$ and have no poles.

\vskip 1cm
\section{The nonlinear liquid drop model}
\vskip 1cm

The liquid drop model has an infinite-dimensional 
Hamiltonian structure described by a nonlinear Hamiltonian function. 
Such systems can be treated in the same way as the
finite-dimensional ones, excepting some difficulties related with the
differentiability of the flow and with the definition of the symplectic
structure, [12]. These occure, on one side,  due to the fact
that the vector fields are only densely defined (since we are dealing with
partial differential equations) and, on the other side, due to the fact
that the linear Hamiltonian and the nonlinear one could have different 
symplectic structures (different associated symplectic forms, phase spaces,
etc.). If these difficulties are overcome, the
theory of Hamilton equations and the invariants and conservation laws 
may be used, too.

The dynamics governing traveling waves and small (even
anharmonic) oscillations in a perfect irrotational fluid,
in 1+1 space-time dimensions, 
can be described by a scalar field $\Phi (x,t)$ and a 
Hamiltonian $H_{[\Phi ]}=\int _{D}h dx$ with the Hamiltonian density $h$
\begin{eqnarray}
H_{[\Phi ]}=\int _{x_1 }^{x_2 } \biggl ( {1 \over 2} \Phi _{t}^{2}
-{{1} \over {V^2 }}|\Phi _x |^2 +F(\Phi ) \biggr ) dx
\end{eqnarray}
where $x_{1,2}=\pm \infty $ for the flow along the line ($D=R$),
and $x_{1,2}=0,\pi $ ($x\rightarrow \phi \in D=[0,\pi ]$),
for a flow on a closed manifold. 
The first term in the integrand stays
for the kinetic energy ($T$) and the next two for the potential energy ($U$), 
the last one describing the restoring forces in a general manner. Such potentials occur
in quantum theory of self-interacting mesons where the function $F$ governs 
the nonlinear part of the interaction. The configuration space for this problem
is some space of smooth real fields with the associated symplectic form
generated by the usual $L^2 $ inner product.
This approach can also provide a Lagrangian $L=T-V=\int _{D}{\cal L} dx$ 
and the equation of motion is
\begin{eqnarray}
\Phi _{t}=\Phi _{tt}-V^2 
\Phi _{xx}-\partial _{x}{{dF} \over {d\Phi }}.
\end{eqnarray}
This formalism provides periodical harmonic solutions in $x$
for $F=0$ and nonlinear waves for general $F$.
In order to
obtain localised solutions, we have to use another type of Hamiltonian, i.e.
a KdV (or modified KdV, etc.) one. This could be provided by the Hamiltonian
\begin{eqnarray}
H[\Phi ]=\int _{D}h(\Phi,\Phi _x ,\Phi _{xx}, ...)dx,
\end{eqnarray}
where $D$ is again the domain of the generic space
coordinate $x$, [9]. The Hamiltonian vector field id given by
$X_{H[\Phi ]}={{\partial } \over {\partial x}}{{\delta h} \over {\delta \Phi
}}$,
where $\delta /\delta \Phi $ is the functional derivative of $h$ 
with respect to the $\Phi $
\begin{eqnarray}
{{\delta h} \over {\delta \Phi }}=
\sum_{k \ge 0}(-1)^{k+j}\partial _{t}^{j}\partial _{\phi }^{k}
{{\partial h}\over {\partial \Phi _{\phi }^{k}\Phi _{t}^{j}}},
\end{eqnarray}
with $\partial _{t}^{j}=\partial ^{j}/ \partial t^{j}$,
$\Phi _{\phi }^{k}=\partial ^{k}\Phi / \partial {\phi }^{k}$, etc.
The corresponding Hamilton equation becomes
\begin{eqnarray}
{{\partial \Phi } \over {\partial t}}=
-{{\partial } \over {\partial x}}{{\delta h} \over {\delta \Phi }},
\end{eqnarray}
i.e. $\Phi (x,t)$ are the integral curves of $X_{H_{[\Phi ]}}$.
Localised solutions must vanish at infinity for $D=R$, and must be periodical
or rapidly decreasing functions for $D=[0,\pi]$. A symplectic structure
on these spaces of functions is given by the skew-scalar product, [12] 
of two arbitrary functions $\Phi _1 $ and $\Phi _2 $
\begin{eqnarray}
\omega (\Phi _1 ,\Phi _2 )={1 \over 2}\int _{D}\biggl (
\int_{x_2}^{x}(\Phi _1 (x)\Phi _2 (y) - \Phi _1 (y) \Phi _2 (x) )
dy\biggr ) dx.
\end{eqnarray}
However it is a  qualitative difference between the two Hamiltonians,
eqs.(36,38), occuring from their different symplectic geometries. This difference
has consequences in the existence of two different phase
spaces, and finaly, in the difficulty of smoothly connecting 
their characteristic solutions, i.e. linear
oscillations and solitary waves.

The liquid drop model consists in the sum of the 
kinetic and potential energy of the fluid, $E=T+U$.
All the terms in $E$ depend on two functions: the shape of the surface
$\xi (\theta ,\phi ,t)$ and the potential  flow $\Phi (r, \theta ,
\phi ,t)$. In the following we use for the shape function the factorization
given in eq.(2) and we consider only one canonical coordinate, $\phi $.
In the following we consider only those solutions $\xi (\theta ,\phi ,t)$ 
in which
the coordinate $\theta $ describing the transversal profile of the
traveling wave is considered to be decoupled of $\phi $, eq.(2).
All the terms dependent of $\theta $ will be absorbed in the 
coefficients of some integrals, and will become the parameters of the model. 
In this case the energy becomes a functional of $\eta $ only.

The potential energy $U$ has at least three terms:
the surface energy ($U_S$), the Coulombian energy ($E_C$) and the shell
energy ($E_{sh}$). The first term describing the surface contribution is
\begin{eqnarray}
U_S =\sigma ({\cal A}_{\xi }-{\cal A}_{0}),
\end{eqnarray}
where $\sigma $ is the surface pressure coefficient,
${\cal A}_{\xi }$ being the area
of the deformed nucleus and ${\cal A}_0$  the area of the
spherical nucleus, of radius $R_0$. 
Both these areas encircle the same volume $V_0=
{{4\pi {R}_{0}^{3}} \over 3}$ of the
nucleus. The surface of the deformed nucleus 
depends only of the shape function
$\xi (\theta ,\phi ,t)$, eqs.(1,2,23).
It is possible to expand eq.(23) for small deformations $\xi /R_0<<1$,
in its Taylor series in all its three arguments $\xi, \xi _{\theta },
\xi _{\phi }$, around 0. By introducing the shape function,eq.(2)
in eq.(23) we obtain, in the third order
$$
U_{S}[\eta ]=\sigma R_{0}^{2}\int_{0}^{2\pi }\biggl [
2S_{1,0}^{1}\eta +(S_{1,0}^{1}+{1 \over 2}S_{0,1}^{1}){\eta }^2
$$
\begin{eqnarray}
+{1 \over 2}S_{1,2}^{1}{\eta }^3
+{1 \over 2}S_{2,0}^{-1}{\eta }_{\phi }^{2}\biggr ]d\phi 
+{\cal O}_{3}(\eta ,\eta _{\phi }),
\end{eqnarray}
where ${\cal O}_3(\eta ,\eta _{\phi })$ represents the contributions of 
terms involving higher
orders than 3 in the function $\xi $ and its derivatives.
We introduce the notations
\begin{eqnarray}
S_{n,m}^{k}=\int_{0}^{\pi }g^{n}g_{\theta }^{m}{\sin }^{k}\theta 
d\theta .
\end{eqnarray}
The term proportional with $\eta \eta _{\phi }^{2}$ 
in eq.(23) is neglected 
because it belongs to the 
fourth order, since the second derivative to the square introduced a 
factor of $\eta _{0}^{3}/L^2 $
which is smaller compared with $\eta _{0}^{3}$, 
($\eta _{0}$ being the amplitude of the perturbation).

The volume conservation condition
\begin{eqnarray}
V_0=\int _{\Sigma }{{r^{3}(\theta ,\phi )} \over 3}d\Omega=
{{4\pi } \over 3}R_{0}^{3},
\end{eqnarray}
leads, according with eqs.(1,2) to the restriction
\begin{eqnarray}
\int _{\Sigma }(3\xi +3\xi ^2 +\xi ^3 )d\Omega =
\int_{0}^{2\pi }(3A_{1,0}^{1}\eta +3A_{2,0}^{1}\eta ^2 +A_{3,0}^{1}
\eta ^3 )d\phi =V_{cluster}.
\end{eqnarray}

The second term in the potential energy is given by the Coulomb interaction,
for constant charge density of the nucleus $\rho _{el}$, 
in the volume $V_0 $
\begin{eqnarray}
U_C [\eta ] ={{\rho _e} \over 2}\int _{\cal V}^{'} 
\int _{\cal V} {1 \over {|{\vec r}-{\vec r}'|}}
d{\cal V}d{\cal V}^{'}.
\end{eqnarray}
The Coulomb energy contains three
terms: the self energy of the core of radius $R_0$ and of the bump (the
deformation) and the interaction energy between these. 
Adding these terms, subtracting the Coulomb energy of the initial spherical
configuration and taking into account the terms up to the third order
in $\xi$, we can write the Coulomb energy in the form, [6]
\begin{eqnarray}
U_{C}=U^{(1)}_{C} \biggl [ 1+(C_{1,0}^{1}+C_{1,1}^{1})\int _{0}^{2\pi }\eta
d\phi
+C_{2,0}^{1}\int_ {0}^{2\pi}\eta ^2 d\phi+
C_{3,1}^{1}\int_ {0}^{2\pi}\eta ^3 d\phi
\biggr ]+U_{C}^{(0)},
\end{eqnarray}
where
$$
U_{C}^{(1)}=3.093(Z_0 -Z_{cl})^2 (\rho {\cal V}_{0})^{-1/3},
$$
\begin{eqnarray}
U_{C}^{(0)}=-0.665Z_{0}^{2}(\rho {\cal V}_{0})^{-1/3}+8.6275Z_{cl}^{2}
(\rho {\cal V}_{0})^{-1/3},
\end{eqnarray}
are constants and the Coulombian energy is given in MKS units. The numbers
$Z_0 ,Z_{cl}$ represent the atomic mass numbers of the parent nucleus and
of the corresponding cluster, respectively. The coefficients $C_{ij}$ in
eq.(48) are defined in the form
\begin{eqnarray}
C_{ij}^{k}={1 \over {R_{0}^i}}\int_{0}^{\pi}h^i g^j \sin ^{k} \theta d\theta .
\end{eqnarray}

The shell energy is introduced by considering that the main contribution
to the shell effects in cluster decays is due to the final nucleus, 
close to the double magic nucleus $^{208}$Pb. The spherical core 
$r\le R_0 -h$ represents the final nucleus, which is also
unexcited for the even-even case.
We introduce the shell energy like a measure of the overlap between
the core and the final nucleus, on one side, and between the 
final emitted cluster and the
bump, on the other hand
\begin{eqnarray}
E_{sh}={{ V_{over }}\over {V+[V_0 -(V_{cluster}+V_{layer})]-V_{over}}},
\end{eqnarray}
where $V_{over }$ denotes the volume of the overlap between the volumes of
the initial $V_0$ and final $V$ nuclei, $V_{cluster}$ is the soliton volume,
eq.(46), and $V_{layer}$ is the layer volume on which the soliton is
moving (i.e. $r\in [R_0 -h,R_0 ]$). 
We use this form for the shell energy multiplied with a constant
$U_0$,
choosen such that the total energy of the
system in the state of rezidual nucleus + cluster to be degenerated with the
ground state energy.

The kinetic energy is given by 
\begin{eqnarray}
T={{\rho } \over 2}\int_{\cal V}v^2d{\cal V}=
{{\rho } \over 2}\int_{\cal V}|\nabla \Phi |^2d{\cal V}=
{{\rho } \over 2}\oint_{\Sigma }\Phi \nabla \Phi \cdot d{\vec S},
\end {eqnarray}
where 
$d{\vec S}=R^{2}_{0}(1+\xi _{\theta }^{2}+\xi _{\phi }^{2}
)^{-1/2}(1, -\xi _{\theta }, -\xi _{\phi }
)\sin \theta d\theta d\phi $ is the oriented  surface element 
of $\Sigma $, eq.(1). Explicitly

\begin{eqnarray}
T={{\rho } \over 2}\int_{\Sigma }
{{
\Phi \Phi _r -{{\Phi \Phi _{\theta }\xi _{\theta }} \over {r}}
-{{\Phi \Phi _{\phi }\xi _{\phi } } \over {r\sin \theta }}
} \over {
\sqrt{1+\xi _{\theta }^2 +\xi _{\phi }^2}
}}
dS
\end {eqnarray}
where $dS=R^2 \sin \theta d\theta d\phi $ is the scalar surface element.
Now we use the first boundary condition. By
taking $\Phi _r$ from eq.(16) and introducing it in eq.(53) the dependence
of the kinetic energy of the tangential velocity along $\theta $-direction
$\Phi _{\theta }$
becomes negligible and we obtain 
\begin{eqnarray}
T={{R_{0}^{2}\rho } \over 2}\int_{0}^{\pi}\int_{0}^{2\pi}
{{
R_0 \Phi \eta _t \sin \theta  +{1 \over {R_0}}
\xi _{\phi } \Phi \Phi _{\phi }(1-\sin \theta )} \over 
{\sqrt{1+\xi _{\theta }^{2}+\xi _{\phi }^{2}}}} d\theta d\phi ,
\end{eqnarray}
where we have approximated $r|_{\Sigma }\simeq R_0 $ in the numerator, since
the corresponding higher corrections will occure in higher order than three
in $\xi $.
We use now the second boundary condition, which, together with the first,
gives the form of the functions $f_n$ in the structure of $\Phi |_{\Sigma
}$.
In the following we take the expression of $\Phi $ from eq.(9), 
with its coefficients $f_n$
obtained in eqs.(12,14,18,21) and using the derivatives of $\Phi $
from eqs.(16),
we have for the kinetic energy in eq.(54), in the second order in $\xi $
$$
T[\eta ]={{R_{0}^{6}\rho V} \over {2}}\biggl ( C_{-1,2}^{3}\int_{0}^{2\pi}
\eta _{t} \int_{0 }^{\phi } \eta ^2 ({\tilde {\phi}}) 
d{\tilde {\phi }} d\phi
+2C_{-1,2}^{3}\int_{0}^{2\pi}\eta _{t}  \int_{0 }^{\phi } \eta 
({\tilde {\phi}}) d{\tilde {\phi }}
d\phi \biggr )
$$
\begin{eqnarray}
+{{R_{0}^{7}\rho V^{2}} \over {2}}
\biggl (
(C_{-2,3}^{5}-C_{-2,3}^{6})\int_{0}^{\pi }\eta \eta _{\phi }
 \int_{0 }^{\phi } \eta ({\tilde {\phi}}) d{\tilde {\phi }}\biggr )d\phi ,
\end{eqnarray}
where $C_{ij}^{k}$ are defined in eq.(50) and 
the symbol ${\tilde {\phi }}$ is a dummy variable of integration. Since we are looking only for
traveling wave solutions (which include any sort of oscillations or moving
bumps) we can transform the derivative with respect to time into the
derivative with respect to $\phi $, by using eq.(2), i.e. $\partial _t =
-V \partial _{\phi }$. We perform then an integration by
parts for those term containing the primitives of $\eta $, and we finaly
obtain for the kinetic energy
\begin{eqnarray}
T[\eta ]=T^{(1)}\int_{0}^{2\pi }\eta ^2 d\phi +T^{(2)}\int_{0}^{2\pi }\eta ^3 d\phi ,
\end{eqnarray}
with the parametric functions $T^{(i)}$ given by
$$
T^{(1)} ={{R_{0}^{6}\rho V^2 } \over 2}C_{-1,2}^{3},
$$
\begin{eqnarray}
T^{(2)}={{R_{0}^{6}\rho V^2 } \over 2}(2C_{-1,2}^{3}R_0 +C_{-2,3}^{5}+
R_0 C_{-2,3}^{6}).
\end{eqnarray}

The total energy of the system is the sum of the terms given in
eqs.(43,48,51,56). 
We write it, in the second order, as a functional depending
on the unknown functions $\eta (\phi ,t) ,g(\theta ), h(\theta )$ and the 
free parameter $V$
$$
E[\eta ,g(\theta ),h(\theta )]=
\int_{0}^{2\pi }\biggl [
\biggl ( 2\sigma R_{0}^{2}S_{1,0}^{1}
+U^{(1)}_{C}(C_{1,0}^{1}+C_{1,1}^{1})\biggr ) \eta
$$
$$
+ \biggl ( \sigma R_{0}^{2}\biggl ( S_{1,0}^{1}+{1 \over 2}
S_{0,1}^{1}\biggr ) 
+U_{C}^{(1)}C_{2,0}^{1}+T^{(1)}
\biggr ){\eta }^2
+\biggl (
+{{\sigma R_{0}^{2}} \over 2}S_{1,2}^{1}+U_{C}^{(1)}C_{3,1}^{1}+T^{(2)}
\biggr )\eta ^{3}
$$
\begin{eqnarray}
+{{\sigma R_{0}^{2}} \over 2}S_{2,0}^{-1}{\eta }_{\phi }^{2} \biggr ] d\phi +
E_{sh}[V_{cluster}]+const.
\end{eqnarray}
We note that in the deduction of eq.(58) we have implicitely used the
boundary conditions, in the expression of $T$.
This functional may be interpreted as depending on $\eta $ only, the
function $g(\theta )$ being fixed through the conditions 1-3.
Generally, the  expression $E$ in  eq.(58) 
depends on three functions $\eta (\phi -Vt )$, $g(\theta )$
and $h(\eta )$, in the third
order of approximation.
In order to formulate eq.(58) as the conservation law of the energy,
i.e. the functional $E$ to describe a Hamiltonian functional, we use the
abstract definition of a Hamiltonian system as the Lie algebra of the
Poisson bracket
\begin{eqnarray}
\{ F_1 ,F_2 \} =\int _{0}^{\pi }\int _{0}^{2 \pi }\biggl (
{{\delta F_1 }\over {\delta \xi }}{{\partial }\over {\partial \phi }}
{{\delta F_2 }\over {\delta \xi}}+
{{\delta F_1 }\over {\delta \xi }}{{\partial }\over {\partial \theta }}
{{\delta F_2 }\over {\delta \xi}} 
\biggr )\sin \theta d\theta d\phi ,
\end{eqnarray}
where $F_i =\int \int f_i (\xi ^{k}(\theta ,\phi ,t))\sin \theta d\theta
d\phi $ are generic functionals and $\xi ^{k}$ represents the set of 
the function $\xi $ together with its derivatives with respect to $\theta
,\phi $ up to a certain order [12].
If we take the energy expression, eq.(58) as a Hamiltonian,
$E \rightarrow H[\eta ]$, then the time derivative of any quantity
$F[\eta ]$ is given by 
\begin{eqnarray}
F_t =[F,H].
\end{eqnarray}
We analyse eq.(60) for $\xi (\theta ,\phi ,t)=g(\theta )
\eta (\phi -Vt)$, i.e. the case investigated in
section 2, eq.(2), with respect to the variation of the function $\eta $
only, and for fixed $g$ and $h$ introduced as parametric functions. 
We define $F=\int_{0}^{2\pi} 
\eta (\phi -Vt)d\phi $ and, since $\delta F /\delta \eta =1$,
we get from eqs.(58,60) 
\begin{eqnarray}
{{dF}\over {dt}}=\int_{0}^{2\pi }\eta _t d\phi =\int_{0}^{2\pi}
\partial _{\phi }{{\delta ({\cal A}\eta +{\cal B}\eta ^2 +{\cal C}\eta ^3 +
{\cal D}\eta _{\phi }^{2})}
\over {\delta \eta }}d\phi ,
\end{eqnarray}
where ${\cal A}=2\sigma R_{0}^{2}S_{1,0}^{1}+U_{C}^{1}(C_{1,0}^{1}+
C_{1,1}^{1})$, ${\cal B}=\sigma R_{0}^{2}(S_{1,0}^{1}+{1 \over 2}
S_{0,1}^{1})+U_{C}^{1}C_{2,0}^{1}+T^{(1)}$, 
${\cal C}={{\sigma R_{0}^{2}}\over {2}}S_{1,2}^{1}
+U_{C}^{(1)}C_{3,1}^{1}+T^{(2)}$ 
and ${\cal D}={{\sigma R_{0}^{2}}\over 2}S_{2,0}^{-1}$, from eqs.(58).
Eq.(61) gives
\begin{eqnarray}
\int_{0}^{2\pi }\eta _t d\phi =\int_{0}^{2\pi }
(2{\cal B}\eta_ {\phi }+6{\cal C}\eta \eta _{\phi }-2{\cal D}
\eta _{\phi \phi \phi })d\phi .
\end{eqnarray}
Eq.(62) lead to the KdV equation, similar with the Euler approach in
section 2.
Therefore we have shown that the energy of the nonlinear liquid drop model,
eq.(58), can be interpreted as the Hamiltonian of a 
of the one-dimensional
KdV equation [5,7-10,13,14],
in agreement with the result obtained in section 2, eq.(30), from the Euler
equation approach.
The coefficients of the terms in the above expression depend on two arbitrary functions
but of different argument, and hence they represent a parametric dependence
and are not involved in the Hamiltonian dynamics of the function $\eta (\phi
,t)$ describing the traveling wave profile in the direction of propagation.
Similar de-coupling of the coordinates is used in the theory of 2-dimensional 
solitary wave as in the Kortweg-Petviashvili (KP) equation [12,13].  

\vskip 1cm
\section{The cnoidal and solitary wave solutions}
\vskip 1cm

The KdV equation in its most general form 
\begin{eqnarray}
A\eta _t +B\eta _{\phi } +C\eta \eta _{\phi }+D\eta _{\phi \phi \phi }=0,
\end{eqnarray}
has three classes of exact solutions, 
depending on the three initial conditions, for
the same values of its coefficients, $A,B,C,D$.
The most general steady-state solution of the KdV eqution has the form
of an oscillation that can reduce to the simple-pulse solution in the limit
that the oscillations period tends to infinity.
This solutions can be classified  using the phase space portrait, Fig.1, in 
cnoidal waves, solitary waves and singular solutions, Fig.2. 
In the following we denote with $V$ the
traveling velocity (phase velocity), with $L$ the half-width
and with $\eta _{0}$ the amplitude of the solutions, eqs.(32,33).
For small amplitudes $|\eta _0 |\ll 1$, i.e. $C\simeq 0$, the KdV 
equation with has approximate solutions in the form of 
stationary small amplitude
oscillations, if $sign(CD)=+1$. 
This is the case when the nonlinear term (the coefficient $C$ in eq.(31))
is neglected, the dispersion law becomes $L=2\pi / V^{3/2}$ and the group velocity 
is
$V_{gr}=3V$. These solutions are instable and fall into cnoidal waves, if
their amplitude is increased with the increasing of the energy.

The algebraic
convenience is obtained by writing the steady state solutions in the form of
eq.(2) with $\eta _{t}=-V\eta _{\phi }$. This allows us to use,
in the following, the variable $z=\phi -Vt$. The KdV equation
reduces to an ordinary differential equation for which a first quadrature
may be immediatley obtained. A second quadrature is possible, after the
first one, by multiplication of the resulting equation with $\eta _{z}$.
We find that
\begin{eqnarray}
\eta _{z}^{2}={{2C}\over {3D}}\eta ^3 +{{AV-B}\over {D}}\eta ^2+ a\eta +b,
\end{eqnarray}
where $a,b$ are constants of integration. Eq.(64) factorizes into
\begin{eqnarray}
\eta _{z}^{2}=-4(\eta -\alpha _1 )(\eta -\alpha _2 )(\eta -\alpha _3
)=-U[\eta ],
\end{eqnarray}
with $\alpha _1 \le \alpha _2 \le \alpha _3 $ real roots and with the same
interpretation for $U[\eta ]$ like in eq.(58) and Figs.1, 
of the kinetic (LHS of eq.(62)), 
potential ($U$) and total ($E$) energies.
Since $\eta _z $ is real, for finite amplitude oscillations $\eta $ must be
confined to the range $\alpha _2 \le \eta \le \alpha _3$, Fig.3.
In between that range, the potential energy has a valley which characterizes 
the stability of the solutions. 
Setting
\begin{eqnarray}
\alpha _3 -\eta \biggl ( \sqrt{{(\alpha _3 -\alpha _1 )C
}\over {6D}}z\biggr ) =(\alpha _3 -\alpha _2 )w^2
(z),
\end{eqnarray}
we obtain the differential equation
\begin{eqnarray}
w_{z}^{2}=(1-w^2 )(1-k^2 w^2 ),
\end{eqnarray}
with $m^2 ={{\alpha _3 -\alpha _2 }\over {\alpha _3 -\alpha _1 }}$. This
equation defines the Jacobi elliptic functions  $sn(z|m)$ of parameter $m$ 
of amplitude $z$ [15]
\begin{eqnarray}
sn(z|m)=\sin \beta , \ \ \ 
z=\int_{0}^{\beta }{{dx}\over {\sqrt{1-m\sin ^2 x}}}, 
\end{eqnarray}
and evidently $z(\pi /2,m)=K(m), cn^2 (z|m)=1-sn(z|m)$.
The square of the cnoidal sinus and cosinus
oscillates between 0 and 1, when $m$ takes real values in $[0,1]$, with a
period equal with $2K(m)$, where
\begin{eqnarray}
K(m)=\int_{0}^{1}{{dx}\over {[(1-x^2 )(1-m^2 x^2 )]^{1/2}}},
\end{eqnarray}
represents the Jacobi elliptic integral.
The solutions become
\begin{eqnarray}
\eta (\phi -Vt)=\alpha _2 +(\alpha _3 -\alpha _2 )cn^2 \biggl ( \sqrt{{{C(\alpha _3
-\alpha _1 )}\over {6D}}}(\phi -Vt)\bigg | m \biggr ) .
\end{eqnarray}
The solution $\eta $ oscillates between $\alpha _2$ and $\alpha _3$, with a period
$T=2K(m)\sqrt{{6D}\over {(\alpha _3 -\alpha _1 )C}}$, as one can see from the
negative potential valley in Fig.3. This solution is not allowed to exist
outside of this valley. However, the shape of the valley can be
modified by choosing different values for the parameters $\alpha _i $, 
i.e. one can associate different potential pictures to different sets of
initial conditions. The function $cn(z|m)$
has two limiting forms $cn(z|0)=\cos z$, $cn(z|1)=$sech $z$, see Fig.2, too. 
The potential for both these limits
is presented in Fig.3.
If $\alpha _2 $ approaches $\alpha _1 $, $m$ approaches 1 and $T,K
\rightarrow \infty$. We obtain in this limit a localised solution
\begin{eqnarray}
\eta (\phi -Vt)=\alpha _2 +(\alpha _3 -\alpha _2)cn^2 \biggl (
\sqrt{{{C(\alpha _3 -\alpha _2 )}\over {6D}}}(\phi -Vt)\bigg | 1
\biggr ),
\end{eqnarray}
which is a bump of half-width 
$L=\sqrt{{6D} \over {C(\alpha _3 -\alpha _2 )}}$
and amplitude $\alpha _3 -\alpha _2 $ above a reference level $\alpha _2$.
More, since $\alpha _1 +\alpha _2 +\alpha _3 ={{3(B-AV)}\over {2C}}$, we can
write, in the limit $\alpha _2 =0$ (so that $\alpha _3 ={{3(B-AV)} \over {2C}}$)
\begin{eqnarray}
\eta (\phi -Vt)=\eta _0 sech ^2 \biggl [
\sqrt{{{\eta _0 C}\over {12 D}}}(\phi -Vt)
\biggr ],
\end{eqnarray}
which is exactly the soliton solution, i.e. eq.(32). From Fig.3 we see
that the solution is defined  between the limits $\alpha _1 ,\alpha _3 $ 
(the maximal domain of definition) and, due to the
value zero of the derivative of $\eta $ in $\alpha _1$ (the saddle point), 
it requests infinite
time to approach this limit, therefore it has a high stability. 
On the other hand, a
steady-state solution in the form of a small oscillation is obtained when
$\alpha _3 \rightarrow \alpha _2 $ and $m \rightarrow 0 , T\rightarrow \pi
/2$. In this limit we have the solution for small harmonic oscillations 
in the form
\begin{eqnarray}
\eta (\phi -Vt)=\alpha _2 +(\alpha _3 -\alpha _2 )\cos ^2 \biggl (
\sqrt{{{(\alpha _3 -\alpha _1 )C}\over {6D}}}(\phi -Vt)
\biggr ).
\end{eqnarray}
In this latter case the range for $\eta $
becomes very small, in principle a point situated in the bottom
of the potential valley centered on $\alpha _3 $, Fig.3, (but in fact a small domain, since
the KdV equation was deduced in the second order only)
In conclusion, within the same equation, only by modifying the initial
conditions and the parameter $V$, one can obtain both periodic solutions and
localised bumps. A picture showing the smooth deformation from the cosine
function towards the soliton is presented
in Fig.4. Such different choices introduce different potentials, as
in Fig.3. In this way, following a certain path in the space of these
parameters, the dynamical evolution can smoothly approach both limits. 
The solutions depend on 
three free parameters (constants of integration) subjected to two additional
conditions.
On one side we have the condition of the volume conservations, requested by the
physical solutions, eqs.(45,46). On the other side
the solutions must be periodic, i.e. 
\begin{eqnarray}
K\sqrt{{\alpha _3 -\alpha _2 }\over {\alpha _3 -\alpha _1 }}=
{{\pi}\over {n}}\sqrt{\alpha _3 -\alpha _1 },
\end{eqnarray}
for $n=1,2,\dots ,N$ with $N$ always finite ($K(m)$ is bounded from below) 
such that $N\le 2\sqrt{\alpha _3 -\alpha _1 }$. Consequently the whole model
contains only a free parameter which
can be choosen to be one of the $\alpha
$'s, $V$ or $\eta _0 $, and we can draw the potential shape 
$U(r, \eta _0 )$ as a 2-dimensional surface. In Fig.5 we present such 
a contour levels potential picture where
the coordinates are the amplitude of the soliton $\eta _0$ (denoted $r$) and
the parameter $\alpha _2 $, (denoted $a$). 
Each cross section of the potential surface
parallel to the $r$-axis represent the potential of the solution, for
a certain initial condition (a fixed value of $\alpha _2 $) and a certain fixed volume.
In the limit of the soliton wave ($r$ around 3)
we have a deep potential hole ($\alpha _2 =\alpha _3 \simeq 1.6$) and in the
limit of the small oscillations ($r$ around $2.4$, $\alpha _2 =\alpha _1 
\simeq 3.2$) 
we obtain again a minimum, embedded into a large potential valley.
The reference level of the soliton ($r\simeq 1.4$) 
is smaller than the average reference
level of the small oscillations, due to the volume conservation.
The evolution of the corresponding solution $\eta (\phi -Vt ;\alpha _2 )$
between these limiting forms is plotted in Fig.6.

\vskip 1cm
\section{The spectroscopic factors}
\vskip 1cm
The spectroscopic factor are defined as the ration between the experimental
decay ($\lambda _{exp}$) over the theoretical one-body constant ($\lambda
_{Gamow}$). For the alpha decay, the relative ratios are well reproduced but
the absolute values are underestimated at least by two orders. Possible
explnations arrise from the fact that the many-body correlations are
neglected in these approaches, the correct antisymmetrization of the channel
wave function, the impossibility of separation in two factors of the
transition element in the integral theory [1,2].
In the present paper we consider that we have introduced the many-body
effects in a phenomenologic way, based on a collective model, resulting in
the possibility of describing of the large enhancement of the alpha cluster
on the nucler surface. hence, we try to give an explanation of the
spectroscopic factors based on this model. By considering the cnoidal waves
as solutions
of both the hydrodynamic equations (Euler) and the liquid drop model
equations, we obtain new shapes which show that, starting from the harmonic
normal modes on the nucler surface and due to the nonlinearities,
some nucleons are grouping together to form the emitted cluster.
In the cnoidal wave description, the spectroscopic factors may be given by
the ratios of the two wave amplitudes in the corresponding wells, evaluated
with the help of barrier penetrability between the two minima, Fig.5,
[4,5].
The spectroscopic factors are given by the penetrabilities of the above
barriers
\begin{eqnarray}
S=exp\biggl (
-{{2}\over {\hbar }}\int_{0}^{\eta _{0}}(A_{cluster}-E[\eta ])^{1/2}d\eta 
\biggr ),
\end{eqnarray}
where $\eta _0$ is the final amplitude of the soliton, Fig.6, .
The function $E[\eta ]$
is the total energy from eq.(57), with the parameter $U_0 $ fitted such that
the second minimum of the potential energy, Fig.7, to be degenerated in
energy with the first one.
The numerical parameters are identical with that one used in the references
[4,5].
The result is compared with similar calculations [4,5,16,17] and with
the experimental preformation factors for $^{208}$Pb, [s]:
$S_{exp}=0.085$, $S_{ref. [4,5]}=0.095$, 
$S_{BW}=0.0063$ and $S_{soliton}=0.07$. The first theoretical 
result was obtained in the frame of a previous
form of the present model [4]. The second result represents the spectroscopic
factor $S_{BW}=(6.3\times 10^{-3})^{{A_{cluster}-1}\over {3}}$,
obtained in [mm] by using a semiempirical heavy ion potential.
However we mention that there exists some ambiguities in the definition of
the spectroscopic factors.

\vskip 1cm
\section{Conclusions}
\vskip 1cm

In this paper we present a nonlinear hydrodynamic model 
describing the new large 
amplitude collective motion in nuclei,  suggested by the new exotic 
alpha (and cluster) decays. 
From the Euler equation, subjected to nonlinear boundary conditions
at the free surface and the inner surface of the fluid layer, we obtain, 
up to the third order in the deviation of the surface from the
spherical shape, a dynamical equation for the radial coordinate
in the form of the Korteweg de Vries eqution.
The new introduced nonlinear terms lead to a 
model describing steady-states as knoidal waves on the surface of the 
nucleus. These solutions are described by the Jacobi elliptic functions and
cover continuously all the range between small harmonic oscillations,
anharmonic oscillations up to solitary wave.  
This model approaches the traditional liquid drop model in
the linear approximation, when the KdV equation transforms into the
Helmholtz eqution and the cnoidal waves approach their periodic limit, i.e.
the normal modes of vibration of nuclei.
From the total energy of the liquid drop
model (surface, Coulombian, kinetic and shell energy), 
calculated in the same order of approximation, it results 
a functional depending on the shape functions which is exactly the
Hamiltonian of the Korteweg de Vries equation, hence a Hamiltonian
formalism applied to the present model lead to
the same dynamical equation.
The final results for the dynamical equations and the energy are in
agreement with the previous (one-dimensional) version  of the same model,
[4,5].
The solitary wave limit of the solutions could describe the preformation of
the clusters on the nuclear surface. We investigate the potential
energy, depending on three parameters: the amplitude of the deformation, 
the phase 
velocity of the
deformation, and the depth of the layer, under the restrictions of constant 
volume of the cluster and periodicity of the solutions 
on the surface. From the expression of the potential energy we 
find a minimum associated with the harmonic vibration,
(small oscillation limit) and an additional minimum corresponding to the 
solitonic shapes as clusters on the nuclear surface.
By choosing the shell effect contribution such that the two minima are
degenerated, we obtain a potential energy profile which could describe
the potential barrier associated with the preformation of such clusters.
The spectroscopic factors, calculated as the ratio of the square amplitudes
in the two minima, are in good agreement with the experiments.

\vfill
\eject

\vfill
\eject
\centerline{\bf {Figure Captions}}
\vskip 1cm
Figs.1
\vskip 0.5cm
The potential picture and the phase space trajectories for some traveling 
wave solutions. We present, in the left side of Fig.1a  the effective cubic 
potential associated with the KdV equation, eq.(30), and in the right side
the corresponding phase space portrait with the curves describing
motions of constant energy. One associates with the finite
potential valley a class of periodic solutions consisting in anharmonic
oscillations (a), centerd around the bottom (A) of the
valley. In their superior limit these solutions are bounded by a special
singular solution (the separatrix - S) of infinite period, hence
a localised bump for the function $\eta (\phi )\simeq \eta (t)$. 
Starting  from the right side of the potential
valley to the saddle point (the origin of the phase space), 
the corrsponding motion represents the soliton solution. The right
limit of the (S) curve gives the soliton amplitude. The amplitude of the anharmonic
oscillations (cnoidal wave, i.e. periodic solutions of the KdV equation)
are limited by the soliton height. 
Eq.(30) has also other classes of singular solutions, (b-c), which have no
physical semnification for the present model, i.e. not being traveling
waves.

In Fig.1b we present the same picture, for the linear approximation 
of eq.(30). The effective potential is quadratic and gives  
periodic solutions,  harmonic oscillations.  For comparison, we ploted
in the same frame the original cubic potential. One can see how the
nonlinearity developes and creates the pocket and the saddle.

\vskip 1cm
Fig.2
\vskip 0.5cm
We present the three classes of solutions ($\eta (\phi -Vt)$) of the KdV 
equation: the soliton solution (one soliton bump of
height 12 in
the figure), cnoidal waves (large and small  
amplitude anharmonic oscillations ), and the singular (unphysical) solutions
(decreasing to $-\infty $).
The cnoidal waves are similar with some periodic copies of the soliton
shape, and
the small amplitude ones are (in the presented case) just cosine
functions.

\vskip 1cm
Fig.3
\vskip 0.5cm
The effective potential associated with the KdV equation is plotted against
the coordinate $z$, eq.(65). The general
steady-state solution ($cn(z|m)$) has three real zeros, ($\alpha _i $),
which controll the range of the amplitude. It has two limits: the soliton 
($Sech$) and the harmonic oscillations ($cos$). 
The soliton realises the deeper potential valley, where from its stability.

\vskip 1cm
Fig.4
\vskip 0.5cm
The cnoidal cosine function $cn(z|m)$ ploted against its argument and
the parameter $m$. For $m=1$ the function approaches the $Sech$ limit
and for $m=0$ the cnoidal cosine approaches the cos limit. 

\vskip 1cm
Fig.5
\vskip 0.5cm
The contour plots of the potential surface 
$U[\eta ]=U(r,\alpha _2)$ plotted for a fixed volume
$V_{cluster}=4$. One can see the initial state characterized by 
harmonic oscillations, ($r=2.4, \alpha _2 =3.2$), and the final state
of minimum energy associated with the soliton, 
($r=3, \alpha _2 =1.6$). 

\vskip 1cm
Fig.6
\vskip 0.5cm
The shape of the surface associated with the general solution of the KdV
equation, plotted against the coordinate $r=R_0 (1+\eta (\theta ))$ 
and the parameter $V$.
The smooth transition from the harmonic oscillations
form ($V\simeq 0$) to the final solitonic form ($V\simeq 2$), under the volume
conservation restriction, is presented.

\vskip 1cm
Fig.7
\vskip 0.5cm
The potential barrier around a path of minimum energy, 
in the plane of parameters $\eta _0 , \alpha _2 $, from the small
oscillations (first minimum of zero energy) to the solitonic solution
(second minimum). The higher curve represents the energy without
the term $E_{sh}$. The curve which has the second minimum degenerated
in energy with the first one, represents the total energy, with the shell
corrections included. Along the path therer are labeled the corresponding
values of $\eta _0 $.

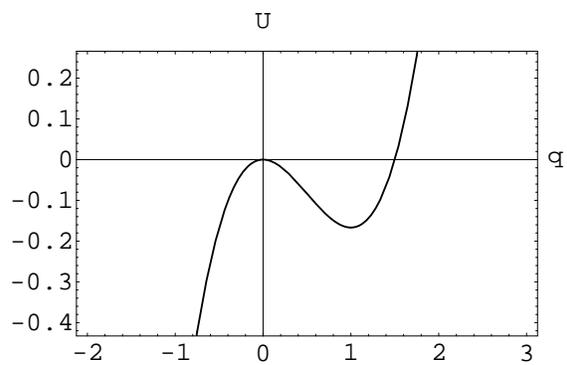
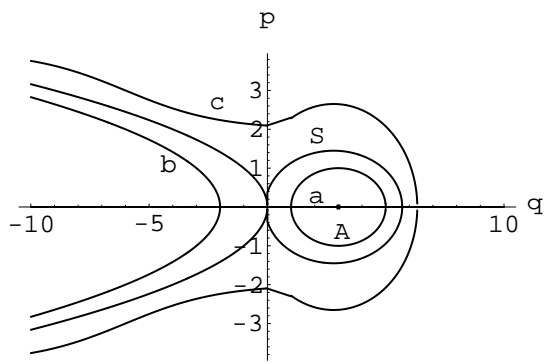
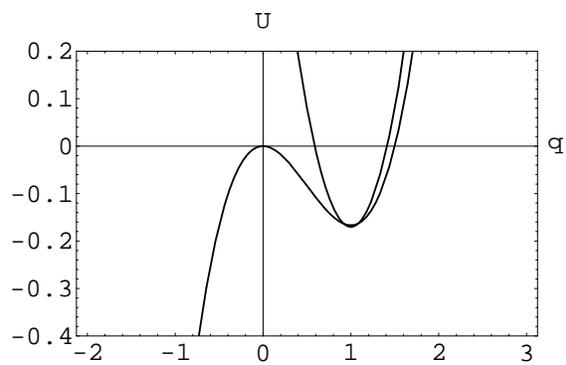
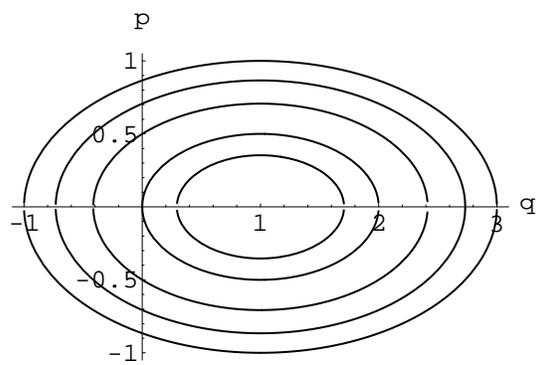
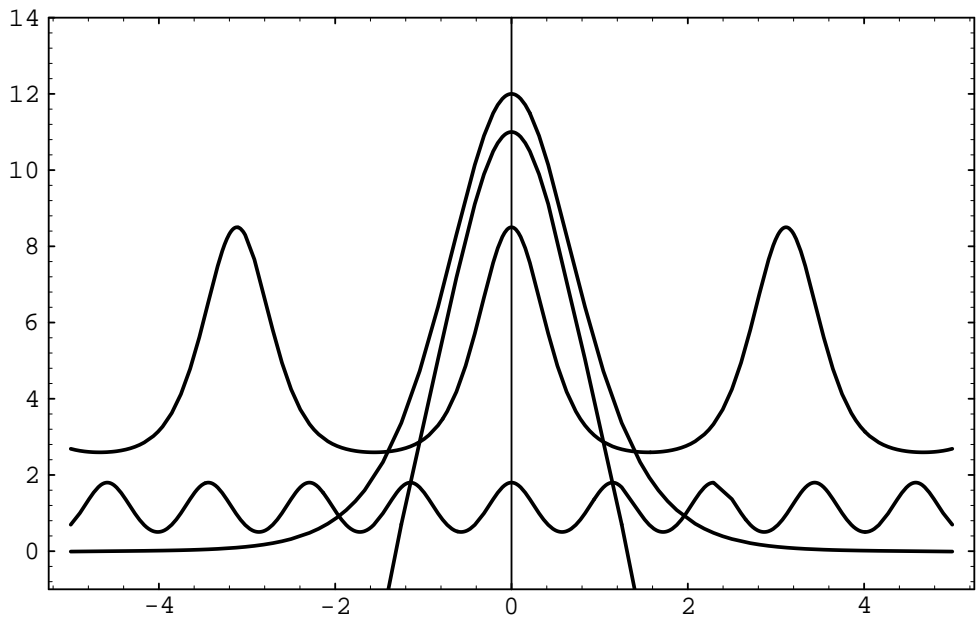
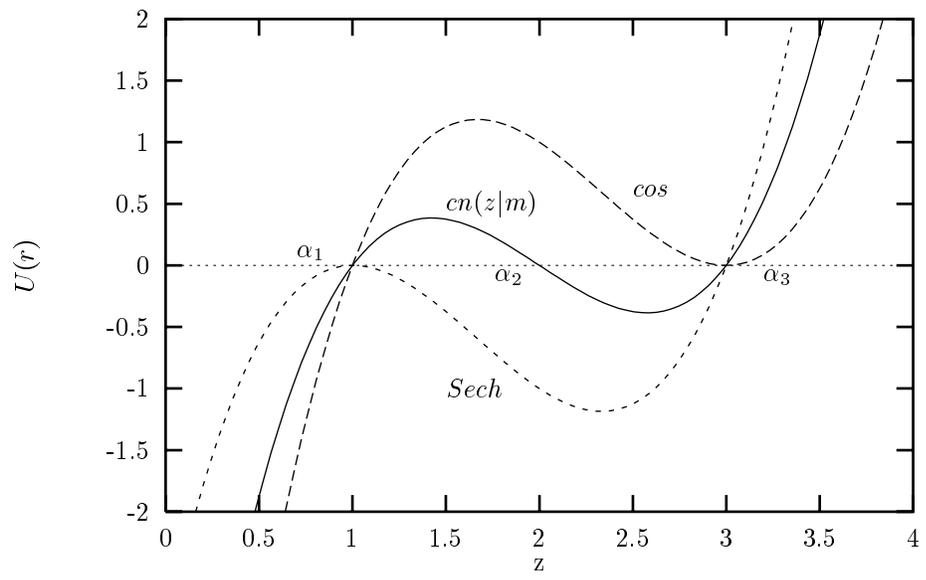
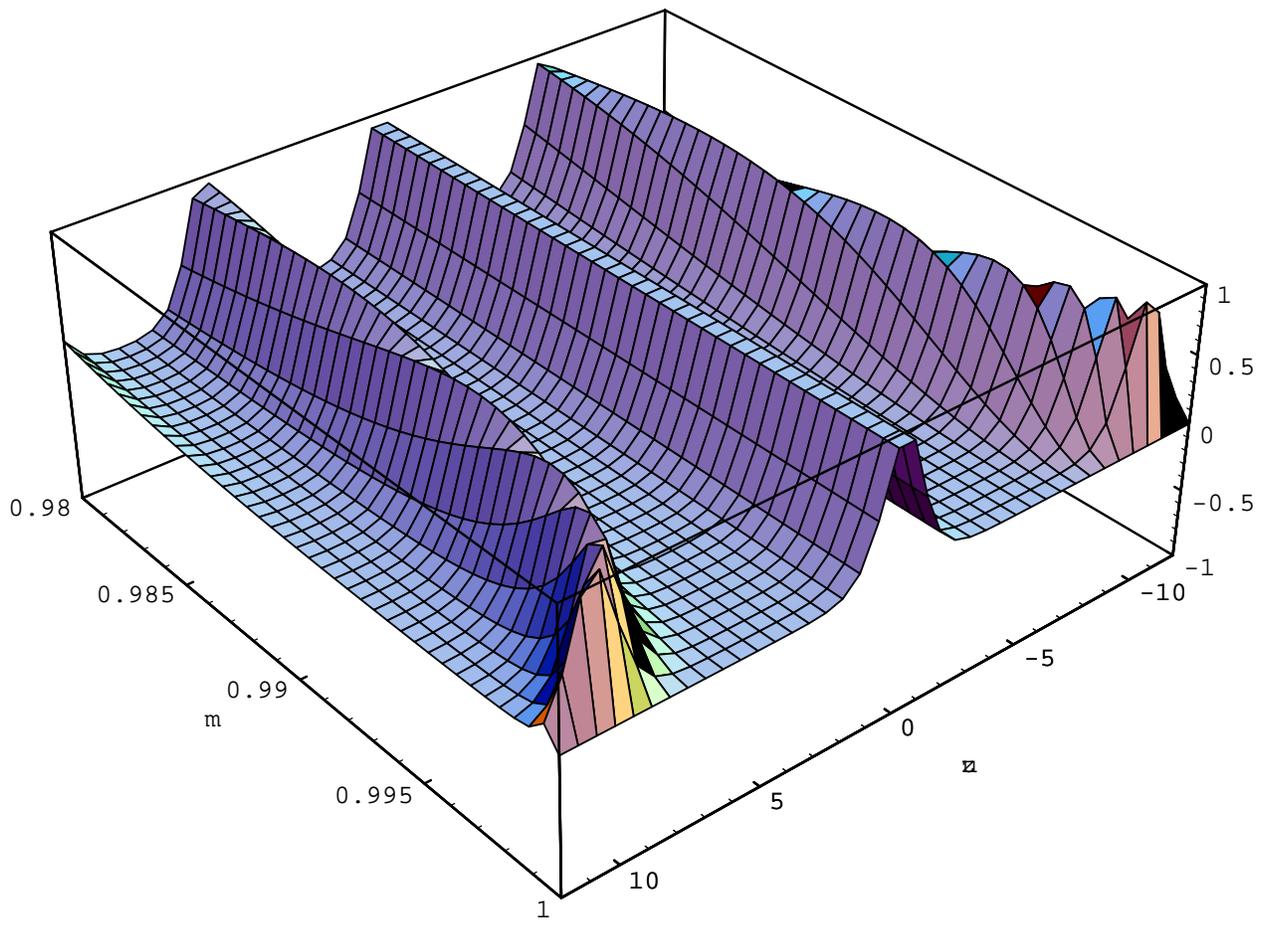
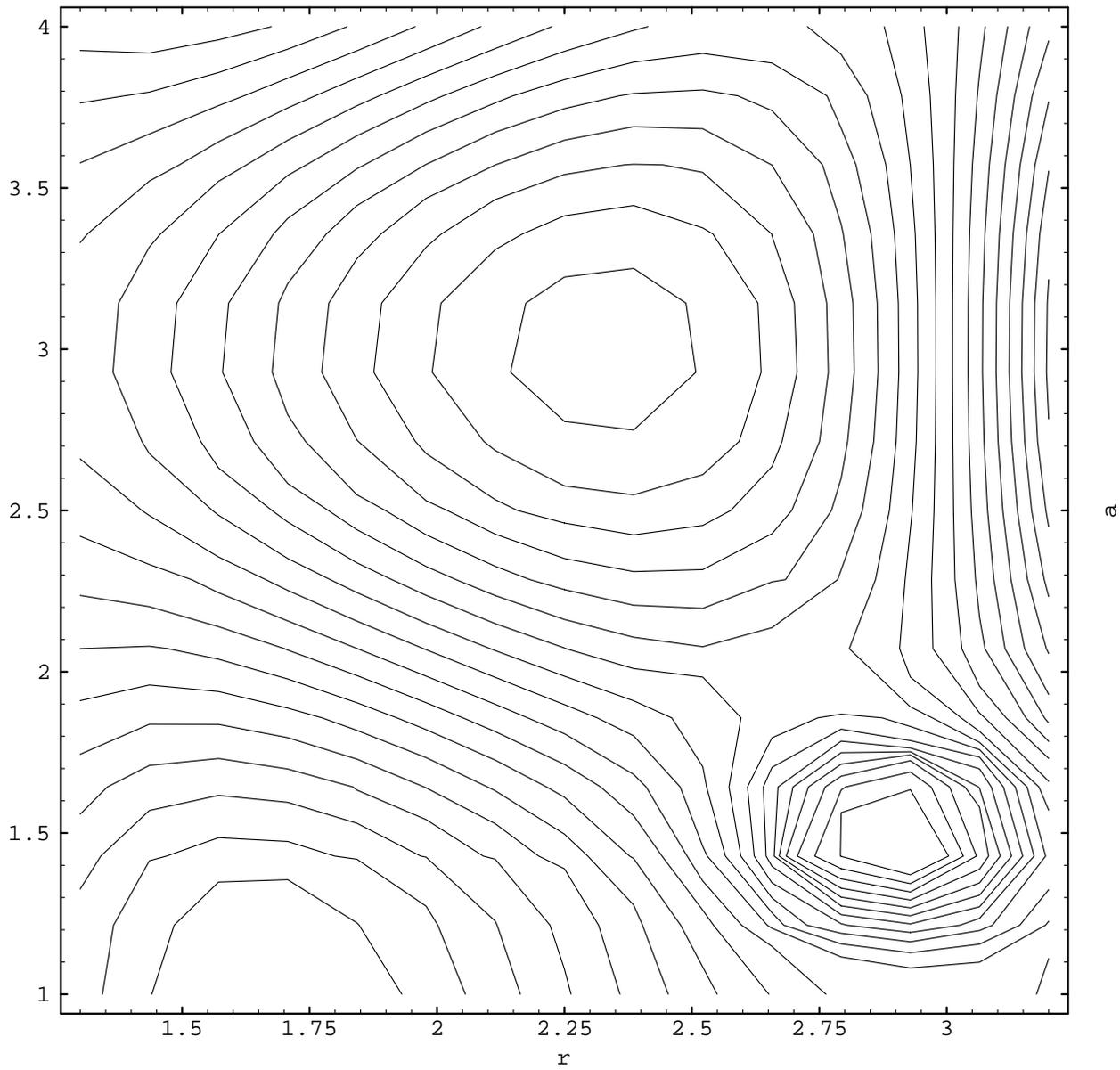
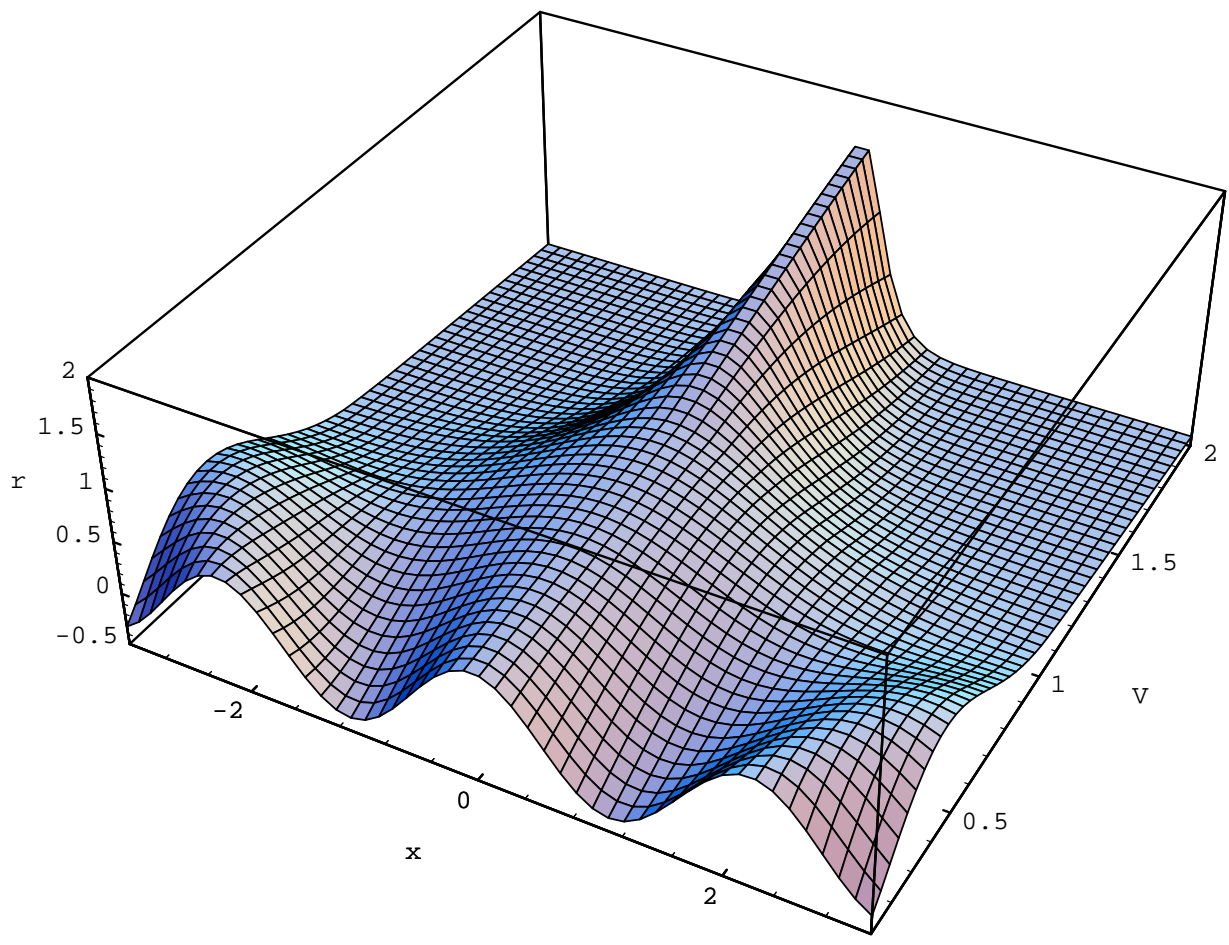
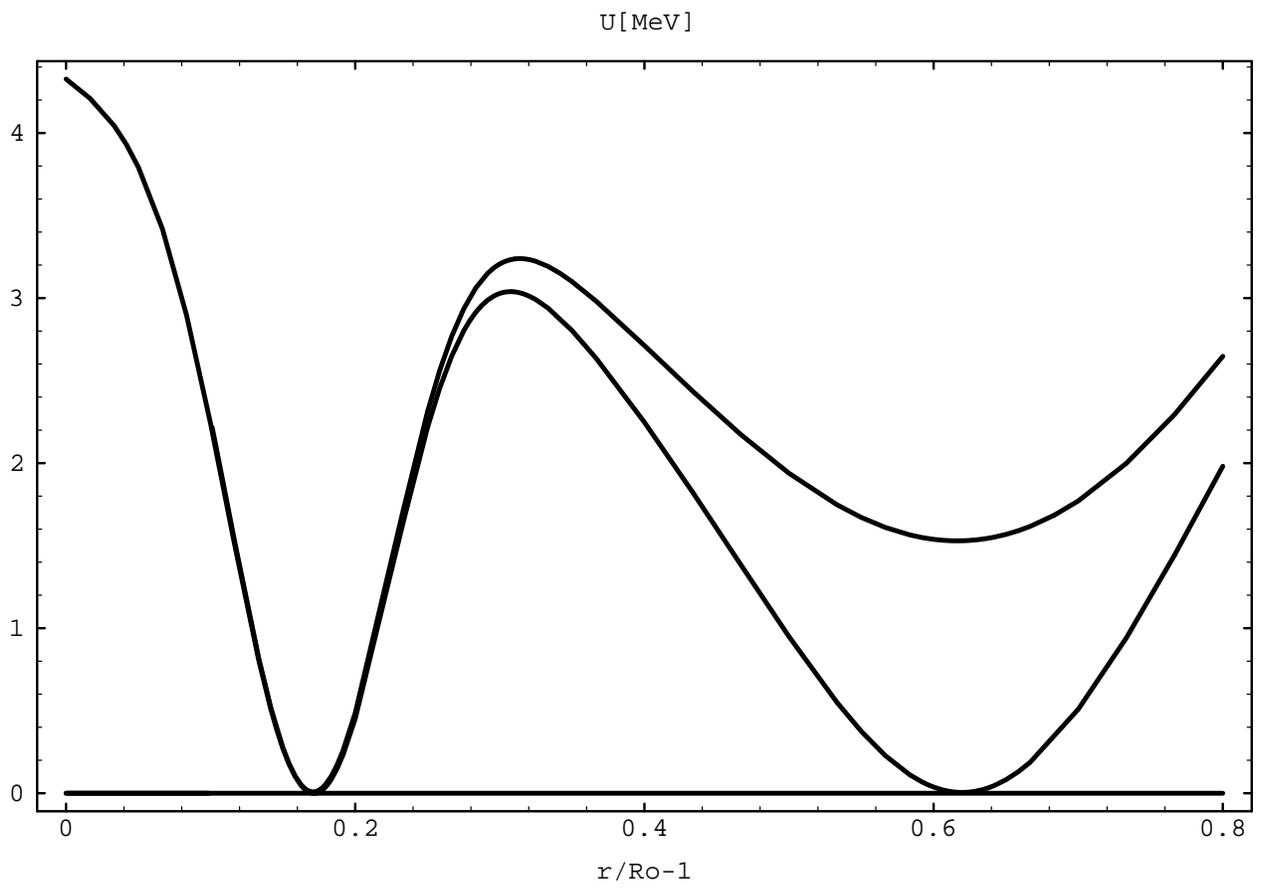
\end{document}